\documentclass{evn2002}
\usepackage{graphicx}
\begin{document}
   \title{Multi-frequency Study of Intraday Variable Sources}

   \author{G. Cim\`o, L. Fuhrmann, T. P. Krichbaum, T. Beckert,
	A. Kraus, A. Witzel
          \and
          J. A. Zensus
          }

   \institute{Max-Planck-Institut f\"ur Radioastronomie, Auf dem
H\"ugel 69, D-53121 Bonn, Germany
         }

   \abstract{
Intraday variability (IDV) of compact extragalactic radio
sources is a complex phenomenon and shows a wavelength dependent mixture
of refractive interstellar scintillation (RISS) (dominant at long
cm-wavelengths) and source-intrinsic effects (dominant at shorter
wavelengths).
Detailed investigations of individual
sources and new high frequency observations suggest a source-intrinsic
contribution to the IDV pattern at least in some sources.
However, the sizes of intraday variable sources at cm-wavelength are
typically smaller than the scattering size set by the ISM in our
galaxy and scintillation must be present, too.

We present new IDV observations in different regimes (from cm
to sub-mm wavelengths) and show how such multi-frequency study can be
used as powerful instrument to describe different aspects of IDV.
   }
   \authorrunning{G. Cim\`o et al.}
   \maketitle
%
%________________________________________________________________

\section{Introduction}

Variability of flat-spectrum quasars on timescales of weeks to years provides
a powerful tool to study the inner regions of these objects.
Variability on shorter timescales, less than one day, was discovered
in the mid eighties (Witzel et al. 1986, Heeschen et
al. 1987). It was found (Quirrenbach et al. 1992) that about 30\,\%
of all compact flat-spectrum sources show such intraday variability (IDV).
The observed rapid variations imply, via the light travel time argument, a
very small source size and a very high apparent brightness temperature
(of up to $10^{21}$\,K), if we consider these variations as intrinsic.

To prevent the Compton catastrophe (which limits the brightness
temperature to $\le 10^{12}\,$K), IDV would
require relativistic boosting with Doppler-factors of up to $D=1000$.
This is much higher than observed with VLBI.
Qian et al. (1991, 1996) proposed a modified shock-in-jet model,
which uses a special geometry to prevent such high Doppler-factors.
In this model, a thin shock moves relativistically along an oscillating jet.
The observed brightness temperature $T^{obs}_B$ then scales with the
$5^{th}$ power of the intrinsic Lorenz factor and only moderate Doppler
factors~($D\le60$) are needed. It, however, appears unlikely to apply for
more than a few special objects, for which furthermore a `suspicious'
geometrical fine tuning between shock thickness and orientation
of the jet axis relative to the line of sight is required (Beckert et
al. 2002).

On the other hand, the sizes of intraday variable sources at
cm-wavelengths are typically
smaller than the scattering size set by the Interstellar Medium (ISM)
in our galaxy.
Hence, IDV sources are small and should show refractive interstellar
scintillation (RISS). While the variations in sources like 0917+62 can
be explained
mainly by RISS (cf. Rickett et al. 1995), this explanation clearly
fails in other sources like e.g. 0716+71 (cf. Wagner \& Witzel, 1995),
where correlated radio-optical IDV is present and most recently IDV at
9\,mm  has been detected (Krichbaum et al 2002).
At least in two sources, the evidence for RISS being the main cause
for rapid variability
has recently become very strong: 0405$-$385 (Jauncey et al. 2000) and
J1819+3845
(Dennett-Thorpe \& de Bruyn 2002). A similar behavior was also
proposed for 0917+624 (Rickett et al. 2001, Jauncey \& Macquart 2001), but
finally was not confirmed by observations with the 100m radio telescope
in Effelsberg (Fuhrmann et al. 2002).
Observations at sub-millimeter wavelengths are important, since
they help to disentangle between extrinsic (dominant at longer cm-wavelengths)
and intrinsic (dominant at mm/optical-bands) causes of the variability.
%
%Alternative models, which are based on coherent emission processes,
%allow brightness temperatures far in excess of the inverse Compton
%limit (Benford 1992, Benford \& Lesch 1998).
%__________________________________________________________________

\section{Observations and Data Reduction}

\subsection{cm-wavelengths:}

All measurements were done with cross scans using the 100\,m radio
telescope of the Max-Planck-Institut f\"ur Radioastronomie (MPIfR) in
Effelsberg.
Our data (Tab.\ref{soulist}) consist of a complete sample
of high declination ($\delta>55\degr$) flat-spectrum
($\alpha>0.5$, we use: $S\propto
\nu^{-\alpha}$) radio sources extracted from the 1\,Jy catalog
(K\"uhr et al. 1981). Here, a new set of observations (carried out in 
March 2000) is included.
Besides the total power, linear polarization information is
collected: the incoming radiation is split in
left- and right components of circular polarization. These signals
enter in a polarimeter that provide as output the Stokes parameters:
I, Q, U.

The data reduction was performed using {\sc cont2}, a task of
the standard software package {\sc toolbox} of the MPIfR. 
Observations of non-variable sources assured a
reliable flux density calibration (accuracy $\sim0.5\,\%$) allowing to
correct for instrumental and atmospheric effects (for details, see
Quirrenbach et al. 1992).
Description of the data and summary of the experiments (since 1989 up
to 1999) are given in Kraus et al. (submitted).
\begin{table}
   \centering
      \caption{List of the complete sample of flat-spectrum sources at high
declination. We also report: galactic latitudes, optical
identifications, known redshifts (Stickel et al. 1994) and the spectral indices
evaluated between 11 and 6\,cm. \textit{Last column}: IDV type (at
6\,cm) from observations (see below in the text).}
         \label{soulist}
\vspace{0.7cm}
	 \begin{tabular}{*{5}{c}|c}
\hline
Name & $b_{II}$~~&ID& z&$\alpha_{11/6}$&Type\\
\hline
0016+73 &  +10.7& QSO&  1.781 &  .16 & I  \\
0153+74 &  +12.4& QSO&  2.338 & -.32 & 0  \\
0212+73 &  +12.0&  BL&  2.367 & -.12 & 0  \\
0454+84 &  +24.7&  BL&  0.122 &  .38 & II \\
0602+67 &  +20.9&  EF&   -    &  .39 & II \\
0615+82 &  +26.0& QSO&  0.710 & -.03 & 0  \\
0716+71 &  +28.0&  BL&   -    &  .21 & II \\
0723+67 &  +28.4& QSO&  0.846 & -.33 & 0  \\
0831+55 &  +36.6& GAL&  0.241 & -.46 & 0  \\
0833+58 &  +36.6& QSO&  2.101 & 1.31 & 0  \\
0836+71 &  +34.4& QSO&  2.172 & -.33 & 0  \\
0850+58 &  +38.9& QSO&  1.322 &  .78 & II \\
0917+62 &  +41.0& QSO&  1.446 &  .25 & 0  \\
0945+66 &  +41.9&  EF&   -    & -.46 & 0  \\
0954+55 &  +47.9& QSO&  0.901 & -.12 & 0  \\
0954+65 &  +43.1&  BL&  0.367 &  .25 & II \\
1031+56 &  +51.9& QSO&  0.459 & -.31 & 0  \\
1039+81 &  +34.7& QSO&  1.254 &  .40 & II \\
1150+81 &  +35.8& QSO&  1.250 & -.09 & I  \\
1418+54 &  +58.3&  BL&  0.152 &  .24 & II \\
1435+63 &  +49.7& QSO&  2.068 & -.23 & I  \\
1637+57 &  +40.4& QSO&  0.750 &  .56 & II \\
1642+69 &  +36.6& QSO&  0.751 & -.26 & II \\
1739+52 &  +31.7& QSO&  1.379 &  .07 & II \\
1749+70 &  +30.7&  BL&  0.770 & -.33 & I  \\
1803+78 &  +29.1&  BL&  0.684 &  .26 & I  \\
1807+69 &  +29.2& GAL&  0.051 & -.35 & II \\
1823+56 &  +26.1&  BL&  0.664 &  .17 & I  \\
1928+73 &  +23.5& QSO&  0.302 & -.01 & 0  \\
1954+51 &  +11.8& QSO&  1.230 & -.14 & II \\
2007+77 &  +22.7&  BL&  0.342 &  .67 & I  \\
2021+61 &  +13.8& GAL&  0.227 &  .10 & 0  \\
\hline
\end{tabular}
 \end{table}
%______________________________________________________________

\section{Statistical analysis}

Statistics provides a basis to investigate the occurrence of
the rapid variations in compact objects. Main instrument for such analysis 
is the \textit{modulation index}:
$$
m[\%]=100\cdot\frac{\sigma_S}{<S>}
$$
(where $<S>$ is the mean flux density and $\sigma_S$ is the rms flux
density variations.)
We define also the \textit{variability amplitude}:
$$
Y[\%]=3\sqrt{m^2-m_{0}^2}
$$
to compare observations taken at different epochs and frequencies
($m_0$ is the modulation index of a non-variable source).

Following Simonetti et al. (1985), we define the first order 
\textit{Structure Function, SF}:
$$
D(\tau)=<(S(t)-S(t-\tau))^2>_t.
$$
This function provides typical timescales and periodicity of
variations and depending on its shape we can define the IDV
type of the objects: a monotonic increase in the SF defines a Type I
source. If the SF shows a maximum then fast
variations are present (Type II). Type 0 denotes non-variable sources.
The last column of tab.\ref{soulist} indicates the IDV types (at 6\,cm) 
in our sample.
One can evaluate the modulation index from the autocorrelation function
as shown in Beckert et al. (this conference). These authors relate
SF to observable parameters assuming some constrains for the ISM and for
the source structure.

\begin{figure}[h!]
\vspace{.8cm}
\hspace{1cm}
   \includegraphics[angle=-90,width=5cm]{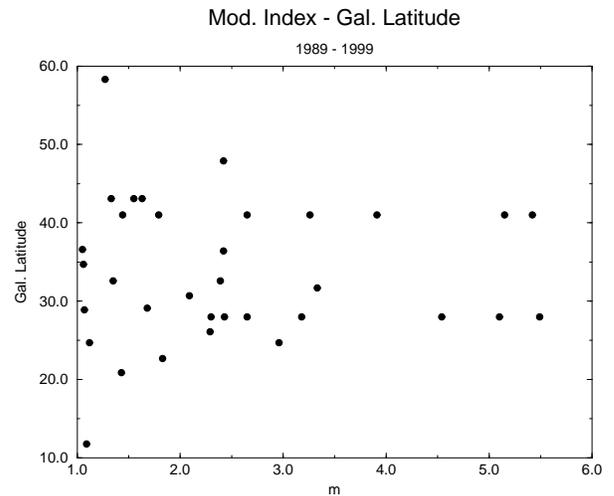}
\vspace{-.7cm}
      \caption{Modulation indices and galactic latitude (summary of
all epochs).
              }
         \label{mb}
   \end{figure}

\begin{figure}[h]
\vspace{1.1cm}
\hspace{1cm}
   \includegraphics[angle=-90,width=5cm]{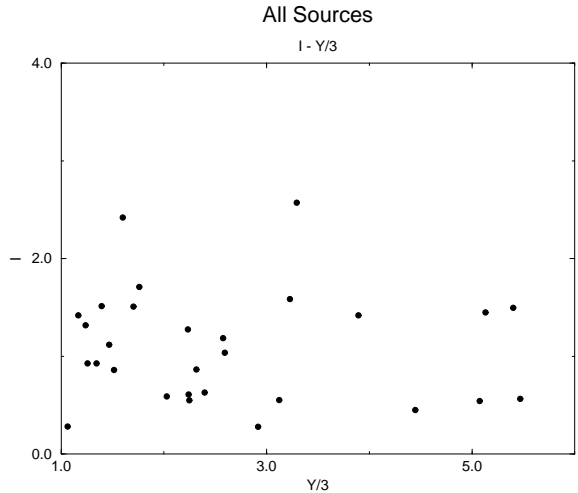}
\vspace{-.7cm}
      \caption{Variability amplitudes at 6\,cm against mean flux
         densities (summary of all epochs).
              }
         \label{alliy}
   \end{figure}

Comparing recent measurements to previous similar data (Quirrenbach
et al. 1992), we immediately see that some objects changed their IDV
characteristics. About one third of the sources in the complete sample
showed variations of type II and overall two thirds can be
considered  type I or type II. However, this fractions are
consistent with previous analysis (Heeschen et al. 1987, Quirrenbach 
et al. 1992). 
Fig.\ref{mb} shows the modulation indices of 
all the variable (type I and II) sources plotted versus galactic
latitude $b^{II}$.
%-------------------------------------------------------------
%-------------------------------------------------------------
(Most of our sources are in the range $20\leq b^{II}\leq 50$.) No
correlation between variations and $b^{II}$ was observed at any
given epoch.
No dependence of the variability amplitudes on the
mean flux density of the sources can be seen, either (fig.\ref{alliy}).
%-------------------------------------------------------------
%-------------------------------------------------------------

\begin{figure}[ht]
\vspace{-0.4cm}
\hspace{-0.3cm}
   \includegraphics[angle=-90,width=9.5cm]{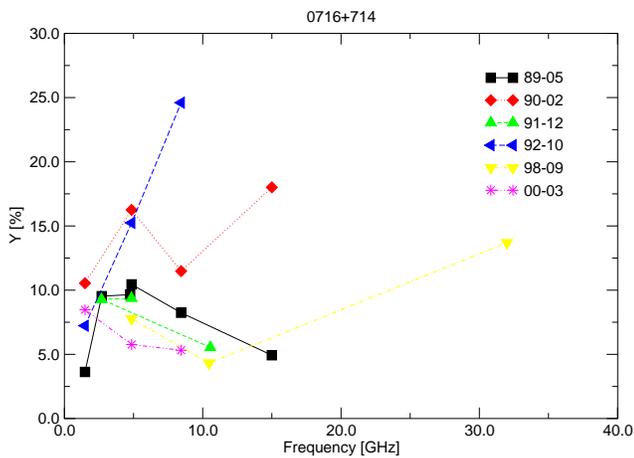}
\vspace{-.7cm}
      \caption{Variability amplitudes against frequency at different
epochs.
              }
         \label{0716nuy}
   \end{figure}

\begin{figure}[h!]
\vspace{-0.2cm}
\hspace{-0.5cm}
   \includegraphics[angle=-90,width=9.5cm]{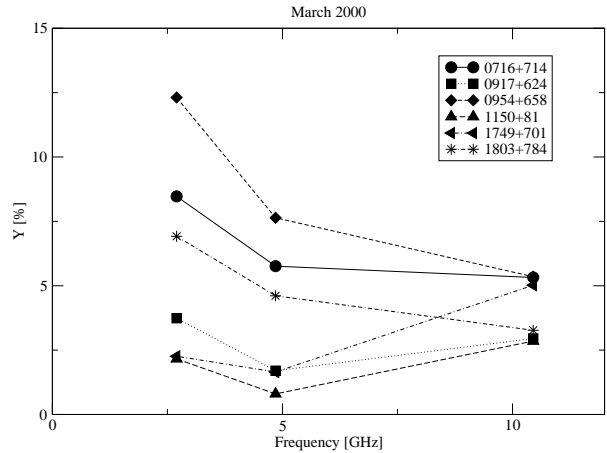}
\vspace{-.5cm}
      \caption{Variability amplitudes against frequency during March
2000 for the variable sources.
              }
         \label{all0003}
   \end{figure}

Scattering theory predicts two regimes of refractive interstellar 
scintillation: weak and strong.
A change between strong
and weak occurs around 3--8\,GHz (Walker 1998): below a critical
frequency the scattering is strong. Such frequency can be determined
observing variations at different frequencies.
Fig.\ref{0716nuy} shows the variability behavior of 0716+714 for 6
different epochs. Immediately we note various patterns indicating
changes either in the source structure or in the interstellar medium.
In some case we see an increase of the variability amplitudes at high
frequency.
%-------------------------------------------------------------
%-------------------------------------------------------------
This is not in agreement with the expectations from RISS,
unless one takes a source intrinsic 
and time variable contribution into account, which increasingly dominates
towards higher frequencies.
Fig.\ref{all0003} shows the behavior of different sources in March
2000. Again, in some sources $Y$ increases with $\nu$.
%-------------------------------------------------------------
%-------------------------------------------------------------

Occurrence of IDV may show dependence
on redshift (considered as compactness indicator). 
Moreover, differences could
occur in time scales or in the magnitude of the variability in BL~Lacs
and Quasars.
Yet, no evidences for such correlations came out of our analysis.
%______________________________________________________________

\section{Sub-Millimeter Observations}

%\subsection{345\,GHz:}
%
In January 2002, a combined total power and polarization
experiment (in collaboration with the MPIfR Bolometer group) 
was carried out at the Heinrich Hertz Sub-Millimeter
Telescope (HHT, a Steward Observatory facility) using the MPIfR
19-channel bolometer array (at 345\,GHz).
Furthermore, in April/May 2002 a radio-sub-millimeter-optical campaign
was carried out. The sub-mm data were taken with the 19-channel
bolometer at the HHT with the on-off technic.
Both set of data were reduced using the program {\sc nic} (which is part of
the Gildas software package and is mainly developed at IRAM
Grenoble).
The data were corrected for the atmosphere opacity ($\tau$) 
using {\sc skydip} scans that measure $\tau_\mathrm{zenith}$.
Strong and compact objects (Ultra Compact H{\sc ii} regions, planetary nebula and
planets) provide a overall calibration accuracy of $\sim10\,\%$.

\begin{figure*}%[h!]
%\vspace{-0.2cm}
%\hspace{-0.5cm}
\centering
   \includegraphics[width=14cm]{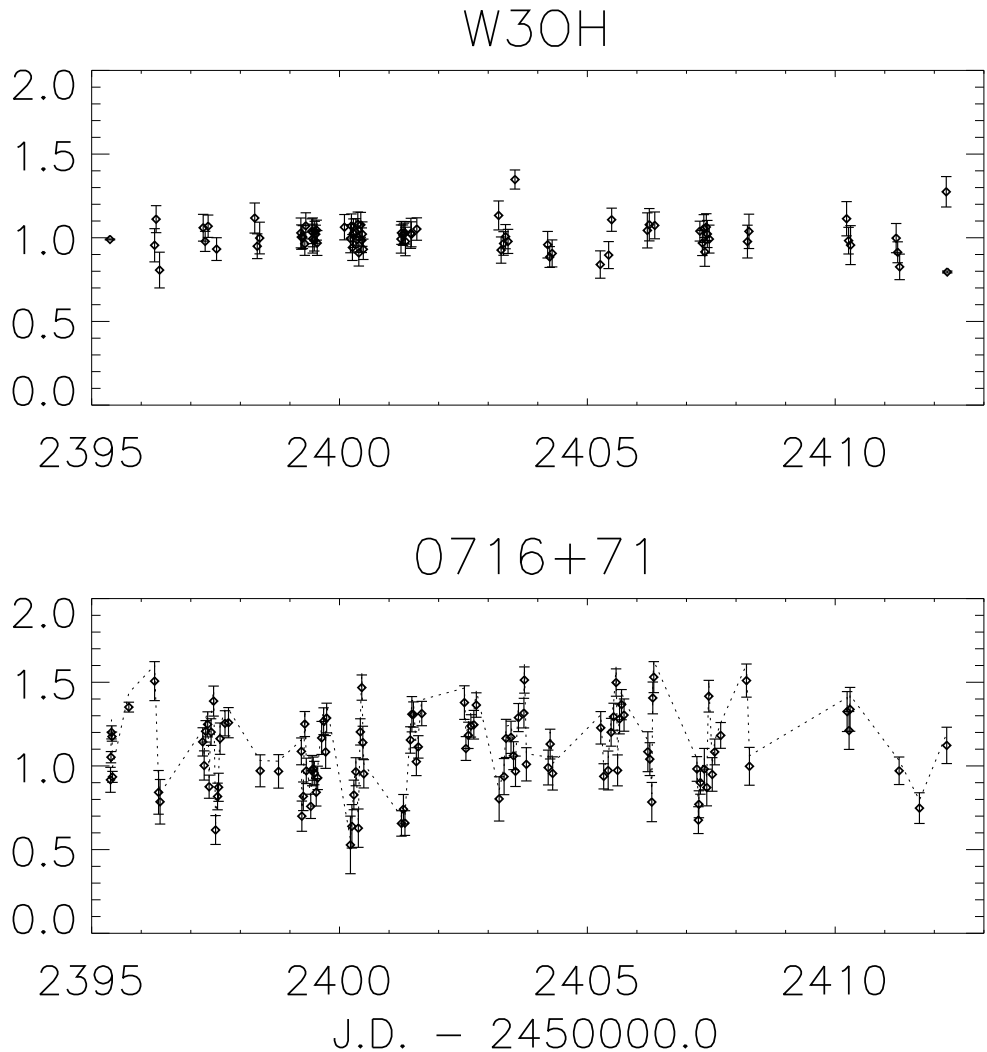}
%\vspace{-.5cm}
      \caption{Preliminary results at 345\,GHz. Tentative detection of
IDV in 0716+714 compared to the calibrator W3(OH).  
              }
         \label{may2002}
   \end{figure*}

%First hint for mm-IDV was shown by Wagner (1998) for 0405$-$385
%using the Swedish ESO Sub-millimeter Telescope in Chile.
A first and encouraging result is the tentative detection of
intraday variations
in the total flux density of the BLLac object 0716+714.
%observed using
%the MPIfR 19-channel bolometer.
For this source, after each scan a calibrator was observed and a {\sc
skydip} was done.
First data analysis indicates 50\,\% 
variations with a time scale $\sim0.5\,$days (fig.\ref{may2002}).
Assuming a lower limit for the redshift of 0.3, such variations imply 
$T_B\simeq 2.2\times 10^{14}\,$K and a Doppler factor, $D\simeq 6$.

Radio-optical correlation seen earlier in 0716+714 and the new
detected sub-millimeter IDV (if confirmed).
will rule out the RISS as sole explanation for the rapid variations in
this source.
%-------------------------------------------------------------
%-------------------------------------------------------------

\section{Conclusions}

Even 15 years after its discovery, 
Intraday Variability in flat-spectrum radio sources remains
a hot and controversial topic in astronomy and still asks for an
explanation.
In the last decade a lot of effort was spent trying to disentangle
the different mechanisms responsible for the observed rapid variations.
The statistical analysis presented here does not give definitive
evidences for one of the proposed models: source-intrinsic or
propagation effects.
Flat spectrum radio sources are strongly variable on
longer time scales. 
%Rising (long term) outbursts are connected with reduced IDV 
%activity ($m$ decreases). 
Fuhrmann et al. (2002) claim strongly quenched
scattering to explain the change in the variability pattern in 0917+62
(Kraus et al. 1999). 
Hence changes in the source morphology and scattering can be
strictly related: variability can show different characteristics at
different epochs due to changes in the apparent source size.
Multi-frequency observations constrain the refractive scintillation
theory and can be combined with VLBI observations
to put some constrains to the ISM structure or the intrinsic source sizes.
%______________________________________________________________

\end{document}